\documentstyle[prl,floats,epsbox,twocolumn,aps]{revtex}

\begin{document}
\draft

\twocolumn[\hsize\textwidth\columnwidth\hsize\csname
@twocolumnfalse\endcsname

\title{\bf The Ground and Low Lying Excited States of
the Two-Dimensional Hubbard Model}
\author{Yoshihiro Asai}
\address{Physical Science Division, Electrotechnical Laboratory (ETL), \\
Agency of Industrial Science and Technology (AIST), \\
Umezono 1-1-4, Tsukuba, Ibaraki 305, Japan}
\date{Received on \today}
\maketitle

\widetext

\begin{abstract}
We have studied the ground state of the two-dimensional (2D) Hubbard
model by using a quantum monte method paying special attention to
the shell structure effect on finite size clusters. Our calculations show 
there is a gap for spin excitations in the ground state and incommensurate
peaks at $(\pi \pm \delta, \pi)$ and $(\pi, \pi \pm \delta)$ in the spin 
correlation function for a low lying excited state. In the ground state,
the long range part of the d-wave superconducting correlation function is
enhanced and the momentum distribution function at the Fermi level
$(\pi,0)$ is rounded. The gap in spin excitations and the momentum
distribution function rounding is consistent with opening a d-wave
superconducting gap in the ground state.
\end{abstract}

]

\narrowtext

The low temperature properties and ground state 
of the doped two-dimensional (2D) Hubbard model have been intensively
studied. This activity increased rapidly after the discovery of the
high-$T_c$ cuprate, because it is widely believed that the single band
model of the 2D $CuO_2$ plane is the essential model for the high-$T_c$
mechanism. Although some novel scenarios have been proposed for the low
temperature state and the ground state of the doped 2D Hubbard model,
little is known with any degree of certainty. One fact which is exactly
known is that no superconducting off-diagonal long range order (ODLRO) 
exists at any finite temperature.~\cite{Tasaki} However it is not known if
there is superconducting ODLRO in the ground state.

The auxiliary field quantum monte carlo methods have been applied
to the doped 2D Hubbard model.~\cite{Reviews} 
The results obtained so far by using the ground state algorithms have not
been sufficient to create a consensus whether or not there is
superconducting ODLRO in the ground state of the model.~\cite{SCs,noSCs}
The difficulty partly comes from the negative sign problem of quantum
monte carlo methods, which prohibits one from studying the ground state of
the doped 2D Hubbard model in wider classes of physical parameter ranges,
carrier densities, band structures, and lattice sizes. Algorithms have
been proposed trying to reduce the negative sign ratio.~\cite{Sign}

We have studied the ground state of the doped 2D Hubbard model
by using the recently developed adaptive sampling quantum monte carlo
(ASQMC) method, with which the negative sign ratio can be greatly 
reduced.~\cite{Asai}
To simulate the Fermi degeneracy among $k$ points around the Fermi
level which may be indispensable to enhance the d-wave superconducting
correlation on the finite size lattices, we have carefully tuned the
non-interacting band structure as well as the carrier density. 
We found keeping the pseudo degeneracy around the Fermi level is quite
essential to our results. This is a difficult requirement for the standard
quantum monte carlo methods.

The Hubbard model we studied has the next-nearest hopping term
and the third nearest hopping term in addition to the nearest
neighbor hopping term:
$
H =
-t \sum_{\langle i,j \rangle \sigma}
( c^\dagger_{i \sigma} c_{j \sigma} + H.C. )
+ U \sum_{i} n _{i \uparrow} n _{i \downarrow}
-t' \sum_{ ( i,j ) \sigma}
( c^\dagger_{i \sigma} c_{j \sigma} + H.C. ) 
-t'' \sum_{ [ i,j ] \sigma}
( c^\dagger_{i \sigma} c_{j \sigma} + H.C. ) ,
$
where the $t$, $t'$, and $t''$ terms are the nearest neighbor
hopping, the next nearest neighbor hopping, and the third
nearest neighbor hopping terms and the $U$ term is the on-site
Coulomb repulsion term. We fixed $t=1$ and it is the unit of energy.
Only the $S_z =0$ subspace was studied.

We have studied the Hubbard model on $6 \times 6$ and $10 \times 10$
lattices. Non-interacting band structure and the filling are tuned 
such that the energy gap between the lowest unoccupied level (LUL) 
close to $(0.4\pi,0.4\pi)$ and the highest occupied level (HOL) at
$(\pi,0)$ : 
$\Delta_{LUL-HOL}$ is less than $0.04$.
The following two cases were studied:
(i) $6 \times 6$ lattice with $t' = -0.1667$, and $t'' = 0.2$. 
The number of electron is $28$, 
(ii) $10 \times 10$ lattice with $t' = -0.2$, and $t'' = 0.05$. 
The number of electron is $84$.
In both cases, we introduced anisotropy $\pm 0.0001$ on the $x$ and $y$
components of $t$ and $t''$.
The band structures obtained with these parameter values may correspond to
those of high-$T_c$ cuprates.~\cite{Tanamoto}
It should be noted that the fillings do not necessarily correspond to
those of real high-$T_c$ cuprate and were chosen for the sake of tuning
$\Delta_{LUL-HOL}$.
We have studied both cases with $U=4$. The $U=6$ case was also studied 
in (i) and some other smaller values of $U$ were adopted to check the
consistency of our results.
We used periodic boundary condition in the $x$ and $y$ directions.

In the case of (i) with $U=4$, we found there are the two quasi-degenerate
states with the lowest energies. These two states were obtained by using
two different trial wavefunctions; the non-interacting wavefunction and
spin polarized Unrestricted Hartree-Fock (UHF) wavefunction. The energy
difference is about $0.01$. Holes are doped around $(\pi,0)$ and its
symmetric points in the ground state and they are doped around
$(0.33\pi,0.33\pi)$ and its  symmetric points in the excited state.
The spin correlation function  
$
< S(\vec{q}) > =
1 / N \sum_{i,j} \exp ( - \vec{q} \cdot (\vec{r}_i - \vec{r}_j) )
< \vec{S}_i \cdot \vec{S}_j >
$
has a broad peak at $\vec{Q} = (\pi, \pi)$
in the ground state but it has incommensurate peaks at 
$(\pi \pm \delta, \pi)$ and $(\pi, \pi \pm \delta)$ in the excited state.

\begin{figure}[htbp]
\begin{center}
\epsfxsize=8.5cm
\epsffile{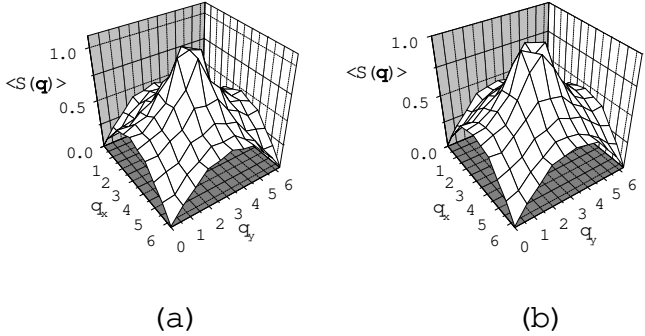}
\end{center}
\vspace*{0.5cm}
\caption{The spin correlation function $< S(\vec{q}) >$ of the ground
state (a) and the excited state (b) of $10 \times 10$ lattice
in the case of (ii) with $U=4$.
The gap is calculated to be $0.04$.}
\label{fig:corrfn}
\end{figure}

\begin{figure}[htbp]
\begin{center}
\epsfxsize=8.5cm
\epsffile{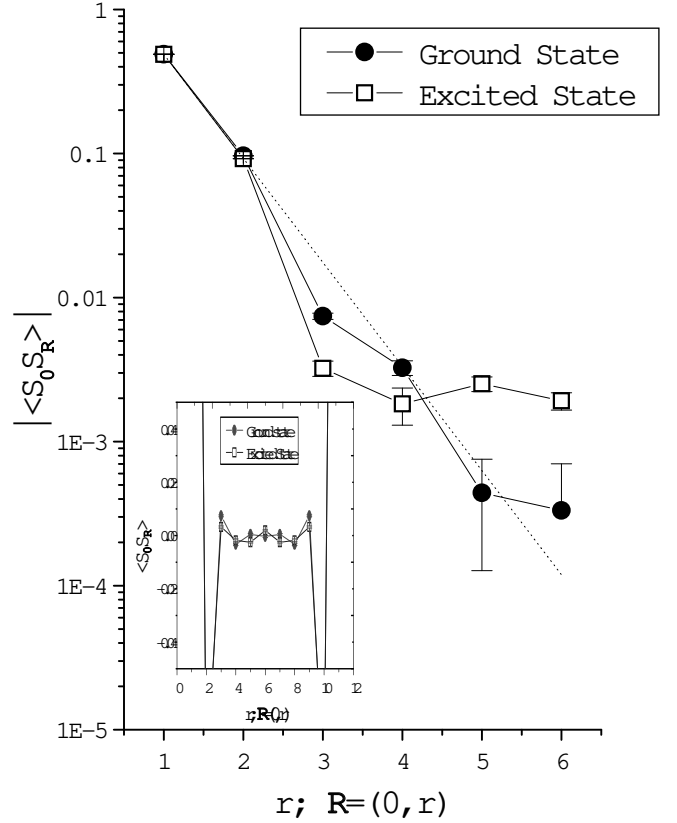}
\end{center}
\vspace*{0.5cm}
\caption{The magnitude of real space spin correlation function
$|< \vec{S}_{\vec{0}} \cdot \vec{S}_{\vec{R}}>|$ in the ground
and the excited states of $10 \times 10$ lattice
in the case of (ii) with $U=4$.
The correlation function is plotted as a function of $r$ where
$\vec{R}=(0,r)$ in the linear-log scale.
The inset of this figure is an original plot of the spin correlation
function $< \vec{S}_{\vec{0}} \cdot \vec{S}_{\vec{R}}>$
as a function of $r$. }
\label{fig:rscorr}
\end{figure}

Similar results were obtained in the case of (ii) with $U=4$.
The energy gap is calculated to be $0.04$.
The spin correlation functions are shown in Fig. \ref{fig:corrfn}.
The broad peak of spin correlation function at $\vec{Q}$ 
suggests strong quantum and/or dynamical
fluctuation induced by the doped hole in the ground state.
The incommensurability $\delta$ in the excited state is calculated to be 
about $0.1\pi$ when the electron density $\rho = 0.84$.
The real space spin correlations
$< \vec{S}_{\vec{0}} \cdot \vec{S}_{\vec{R}}>$ and their magnitude
$|< \vec{S}_{\vec{0}} \cdot \vec{S}_{\vec{R}}>|$
in the ground and the excited state are plotted in 
Fig. \ref{fig:rscorr}.
In the ground state, the magnitude of the spin correlation
decays exponentially, while that of the excited state does not.
This suggests there is a gap in spin excitations in the ground state.

\begin{figure}[htbp]
\begin{center}
\epsfxsize=8.5cm
\epsffile{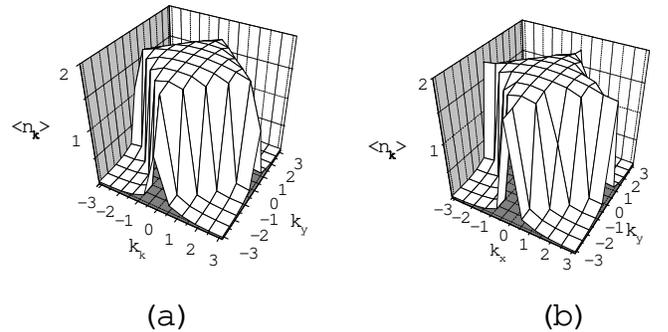}
\end{center}
\vspace*{0.5cm}
\caption{ The momentum distribution function $< n_k >$ of the ground
state (a) and the excited state (b).
The system and the parameter values are the same as the previous
figures.}
\label{fig:nk}
\end{figure}

The momentum distribution function $< n_k >$ of the excited state is 
qualitatively similar to that of the non-interacting model.
It has a sharp drop at the Fermi level.
In the ground state, the drop of $< n_k >$ at the Fermi level $(\pi, 0)$
and $(0, \pi)$ is smeared out and transferred to $(0.4\pi, 0.4\pi)$ and
its symmetric points as shown in Fig. \ref{fig:nk}. $< n_k >$ is therefore 
similar to the momentum distribution function in the d-wave 
superconducting 
state.~\cite{deGennes} 

\begin{figure}[htbp]
\begin{center}
\epsfxsize=8.5cm
\epsffile{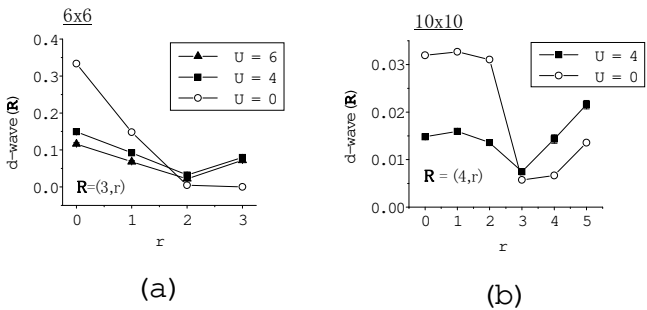}
\end{center}
\vspace*{0.5cm}
\caption{ The d-wave superconducting correlation function
$ d-wave(\vec{R})$
of the ground states of the $6 \times 6$ (a) and the
$10 \times 10$ (b) lattices in the case of (i) and (ii)
as a function of $r$, where $\vec{R} = (x,r)$ .
$U$ is taken as a variable and denoted in the legends of figures.
$x=3$ in (a) and $x=4$ in (b).
Open circles, closed squares, and closed triangles are results
obtained with $U = 0, 4,$ and $6$, respectively.}
\label{fig:d-wave}
\end{figure}

We have calculated the d-wave superconducting correlation function:
d-wave($\vec{R}$) 
$ = <O_{\vec{0}}O^{\dagger}_{\vec{R}}>$, where
$
O^{\dagger}_{\vec{R}} = 
  \frac{1}{\sqrt{2}}
  ( c^{\dagger}_{\vec{R} \uparrow} c^{\dagger}_{\vec{R}+\vec{x} \downarrow}
  - c^{\dagger}_{\vec{R} \downarrow} c^{\dagger}_{\vec{R}+\vec{x} \uparrow} )
  - \frac{1}{\sqrt{2}}
  ( c^{\dagger}_{\vec{R} \uparrow} c^{\dagger}_{\vec{R}+\vec{y} \downarrow}   
  - c^{\dagger}_{\vec{R} \downarrow} c^{\dagger}_{\vec{R}+\vec{y} \uparrow} ) .
$
In the ground states of the cases (i) and (ii), we found enhancement
of d-wave($\vec{R}$) at large distances due to $U$.
d-wave($\vec{R}$) in the ground state of the case (i) is plotted against
$r$ in Fig. \ref{fig:d-wave}, where $\vec{R} = (3,r)$.
The d-wave correlation function of $U=4$ is enhanced against that of $U=0$
at the largest distance. 
We observed the enhancement is persistent up to $U=6$.
d-wave($\vec{R}$) in the ground state of the case (ii) with $U=4$ is plotted
against $r$, where $\vec{R} = (4, r)$ in the same figure.
The d-wave correlation function is enhanced at the two largest
distances. Therefore the enhancement due to $U$ is persistent against
increase of lattice size. The gap in spin excitations and the reduction
of the momentum distribution function at $(\pi,0)$ found in the ground
state are also consistent with a d-wave superconducting ground state.

We have studied the shell structure and/or the band structure
effects on d-wave($\vec{R}$) on the $6 \times 6$ and 
$10 \times 10$ lattices.
The parameter $t'$ in (i) and (ii) is changed as a variable. We put $U=4$.
The enhancement of d-wave($\vec{R}$)
is obtained when $t' \le -0.16$ in the case of (i).
The region shrinks to $-0.2007 \le t' \le -0.1997$ in the case of (ii).
While the width of the flat region of the non-interacting band dispersion 
near $(\pi,0)$ in the case (i) is about $\pi/3$, that of the case (ii) is
almost zero. This may be the primary reason why the enhancement is
achieved in a limited region in the case of (ii). This suggests that the
stability of the d-wave superconducting state is largely influenced by the
non-interacting band dispersion near $(\pi,0)$ and hence $t'/t$ and
$t''/t$.

Implications of our results may be summarized as follows:
(1) The energy difference between the lowest excited state with the 
incommensurate spin correlation and the d-wave superconducting
ground state is very small.
The stabilities of these states are changed by the non-interacting band
structure and the filling. We cannot rule out a possibility that these two
states may coexist in a phase separated region in a high-$T_c$ sample.
This possibility is also a matter of experimental and theoretical
concern today.~\cite{stripes}
(2) In the flat band region near $(\pi, 0)$, doped holes may lose their
phase coherence due to $U$ and may not be able to remain itinerate and
hence they become unstable. To restore the phase coherence, they form
Cooper pairs and become stabilized. This scenario for superconductivity in
the cuprate is consistent with our numerical results and has also been
discussed on the basis of a scaling theory.~\cite{Imada} The reduction of
$<n_k>$ at $(\pi,0)$ due to $U$ is also consistent with this mechanism.

It should be noted that $\Delta_{LUL-HOL}$ implicitly used in most
previous studies, which are often done at closed shell fillings, is
still too large.~\cite{noSCs} Such a situation makes it hard for
superconductivity to have ODLRO in the ground state.
Small $\Delta_{LUL-HOL}$ were adopted by the present author and also
by Kuroki and Aoki (KA).~\cite{SCs} KA studied the small $U/t$ region
($U/t \simeq 1$). In the small $U/t$ region, they observed enhancements
of the d-wave superconducting correlation at large distances.~\cite{SCs} 
The results obtained with larger value of $U/t$ ($U/t \simeq 4$ and $6$)
here may be consistent with their results.

In conclusion, we have studied the 2D Hubbard model with special attention
to the shell structure on the finite size lattice by using the recently
proposed ASQMC method. We found two quasi-degenerate states with the
lowest energies. In the low lying excited state, there are incommensurate
peaks of the spin correlation function at $(\pi \pm \delta, \pi)$ and
$(\pi, \pi \pm \delta)$. In the ground state, there is a gap in spin
excitations and reduction of the momentum distribution function near
$(\pi,0)$. We observed enhancements of the d-wave superconducting
correlation function in the ground state. The gap in spin excitations and
the momentum distribution function rounded at the Fermi level are 
consistent with a d-wave ground state.

The author appreciates Prof. B.A. Friedman for reading this manuscript and
giving some comments. This work is financially supported by the project
E-TK97005 in ETL. Part of the calculations were made by using facilities
of the Supercomputer Center, Institute for Solid State Physics (ISSP),
University of Tokyo and Research Information Processing System (RIPS)
center of AIST, to which the author would like to express thanks.


%
%

%
%
%

\end{document}